\documentclass[a4paper,11pt]{article}
\pdfoutput=1 

\usepackage{jheppub} 
                     
\usepackage{amssymb,amsmath,amsthm,graphicx}
\usepackage{latexsym}
\usepackage{bm}
\usepackage[]{hyperref}
\usepackage{cleveref}

\pdfoutput=1

\begin{document}

\title{To go or not to go with the flow: Hawking radiation at strong coupling}

\author[a,b]{Jorge~E.~Santos}
\emailAdd{jss55@cam.ac.uk}
\affiliation[a]{Department of Applied Mathematics and Theoretical Physics, University of Cambridge, Wilberforce Road, Cambridge, CB3 0WA, UK}
\affiliation[b]{Institute for Advanced Study, Princeton, NJ 08540, USA}
\abstract{We construct the gravitational dual of a one-parameter class of states of strongly coupled $SU(N)$ $\mathcal{N}=4$ SYM at infinite $N$ and asymptotic temperature $T_{\infty}$, on a fixed Schwarzschild black hole background with temperature $T_{\mathrm{BH}}$. The resulting bulk geometry is of the flowing type and allow us to measure Hawking radiation at strong coupling. The outgoing Hawking flux is a function of the dimensionless ratio $\tau\equiv T_{\infty}/T_{\mathrm{BH}}$ and appears to be non-monotonic in $\tau$. At present, we have no field theory understanding for this behaviour.}
\maketitle
\section{Introduction}
\noindent\indent Understanding the behaviour of Quantum Field Theories (QFTs) in curved spacetime is an important problem, not least because we know that the universe does contain regions of very large curvature. A key discovery was Hawking's calculation demonstrating particle production in black hole backgrounds \cite{Hawking:1974rv,Hawking:1974sw}. These particles have a thermal spectrum, confirming that  black holes should properly be thought of as thermodynamic objects. A general argument based on the Euclidean time formalism shows that for any QFT, an equilibrium state on a black hole background (the so-called Hartle-Hawking state) should be thermal \cite{Gibbons:1976es}. However most of what is known about QFTs in curved spacetime comes from calculations involving free or weakly interacting theories. Little is known about the case when the QFT is strongly coupled. In fact, even in the weakly interacting regime there are interesting open puzzles, as for instance those reported in \cite{Akhmedov:2015xwa}.

Gauge/Gravity Duality provides a new way of probing the behaviour of certain strongly coupled QFTs in curved backgrounds. In its most precise and well motivated form, it is the claim that Type IIB Superstring theory on AdS$_{5} \times S^{5}$ is equivalent to $\mathcal{N} = 4$ Super Yang-Mills (SYM) theory on the $(3+1)$ dimensional conformal boundary \cite{Maldacena:1997re,Gubser:1998bc,Witten:1998qj,Aharony:1999ti}.  In the large $N$ strong coupling limit of the boundary gauge theory, the bulk string theory becomes weakly coupled and the string length scale becomes small. In principle this should allow us to study quantum effects in the strongly coupled theory, such as Hawking radiation, by solving classical gravitational equations of motion in the bulk. This technique was explored in \cite{Hubeny:2009ru,Hubeny:2009kz,Hubeny:2009rc,Caldarelli:2011wa}, and reviewed rather beautifully in \cite{Marolf:2013ioa}.

In order to probe the Hawking effect with this technique, we need to take a fixed Schwarzschild boundary geometry. This leads to two classes of bulk gravitational duals: ``Black Funnels" and ``Black Droplets" \cite{Hubeny:2009ru,Hubeny:2009kz,Hubeny:2009rc,Caldarelli:2011wa}. Black Funnels have a connected horizon extending from the boundary black hole into the bulk and out to an asymptotic region. In contrast, the horizon in a Black Droplet solution has two disconnected parts. One extends from and surrounds the boundary black hole (the ``droplet"), and the other is a deformed planar horizon which is not connected to the boundary at all.

Given the fixed black hole geometry on the boundary with temperature $T_{\mathrm{BH}}$, the bulk solutions are characterized by two free parameters: $T_{\infty},$ the temperature of the bulk horizon in the asymptotic region, and $T_{H},$ the temperature of the bulk horizon where it meets the boundary black hole. If $T_{H} \neq T_{\mathrm{BH}}$ then the Euclidean boundary geometry exhibits a conical singularity and the stress tensor diverges at the horizon. These solutions are important (the Boulware vacuum state is described by such a solution \cite{Fischetti:2016oyo}) but we will not consider them in this paper. From this point on we shall always take $T_{H} = T_{BH}.$

Given our remaining freedom in fixing the parameters, there are two special cases of particular interest. The first is the choice of parameters $T_{\infty} = T_{\mathrm{BH}}$. This gives an equilibrium solution and corresponds to the Hartle-Hawking state of the field theory. Physically, a thermal state of the field theory is in equilibrium with a Schwarzschild black hole which emits Hawking radiation at the same temperature. The second case of particular interest is when $T_{\infty} = 0$. This is an out of equilibrium solution corresponding to the so called Unruh state of the field theory. Physically, the field theory approaches its natural vacuum state in the asymptotically flat region (where a natural choice of vacuum state exists) but we now expect to see an outgoing energy flux coming from the Hawking radiation emitted by the higher temperature black hole. This is a good approximation to the state of the field theory that would be obtained after gravitational collapse, but before the resulting black hole has had time to evaporate.

Bulk duals corresponding to the Hartle-Hawking state and to the Unruh state have been constructed previously. For the Hartle-Hawking $T_{\infty} = T_{\mathrm{BH}}$ choice of parameters only black funnel solutions exist \cite{Santos:2012he,Santos:2014yja}. These solutions can be said to describe Hawking radiation in that they contain a black hole in equilibrium with a plasma, but there is no net energy flux. For the Unruh $T_{\infty} = 0$ choice of parameters, the only bulk solutions constructed so far have been of the droplet type \cite{Figueras:2011va}. Surprisingly, this means that they also exhibit no energy flux, at least at the leading order in $N$ in which a classical description of the bulk spacetime is valid. It is expected that Hawking radiation will be present at next leading order in $N,$ but the goal of observing something which can be interpreted as Hawking radiation in a classical solution to the Einstein equation has not yet been realised in the case of the Unruh state.

In this paper we construct for the first time numerical solutions of the black funnel type, containing a Schwarzschild black hole at the boundary, in which $T_{\infty} \neq T_{\mathrm{BH}}$. Such states have been considered previously in the weak coupling context by Frolov and Page in \cite{Frolov:1993fy}. We will study the properties of the system as we vary the ratio $T_{\infty}/T_{\mathrm{BH}}$. We will start with  $T_{\infty}/T_{\mathrm{BH}}>1$ and decrease it passing through $T_{\infty}/T_{\mathrm{BH}}=1$ into the region $T_{\infty}/T_{\mathrm{BH}}<1$. While we are unable to take $T_{\infty}$ all the way to zero, we can lower it significantly, and observe an energy flux corresponding to outgoing Hawking radiation from the black hole. This also means that our bulk horizon is not Killing, where we evade the zeroth law of black hole mechanics \cite{Hawking:1971vc,Hawking:1973uf,Hollands:2006rj} due to the fact that the horizon is non-compact. Other ``flowing funnel" solutions containing two boundary black holes have been obtained previously in the context of AdS boundary geometries \cite{Fischetti:2012ps,Fischetti:2012vt} (see also \cite{Figueras:2012rb,Emparan:2013fha,Sun:2014xoa,Amado:2015uza,Megias:2015tva,Herzog:2016hob,Megias:2016vae,Sonner:2017jcf} for closely related geometries).

The outline of the paper is as follows. Section \ref{sec:cons} explains how to formulate the construction of flowing funnels with arbitrary $T_{\infty}/T_{\mathrm{BH}}$. In section \ref{sec:hor} we detail how to compute some of the horizon properties of flowing geometries, such as its expansion and shear. Section \ref{sec:holo} shows how one can extract the holographic stress energy tensor from the numerical solutions, and in section \ref{sec:res} we present our numerical results. We close with some final discussions in section \ref{sec:dis}.
\section{\label{sec:cons}Constructing the holographic dual}
\noindent\indent We will work in five bulk spacetime dimensions, and thus have a four-dimensional boundary spacetime. We expect that appropriate generalisations exist in higher dimensions. The solution we seek to construct necessarily exhibits temperature gradients across the horizon, the horizon generators cannot be associated with the integral curves of a Killing field \cite{Fischetti:2012vt,Figueras:2012rb}, \emph{i.e.} the horizon is not Killing. So we expect our solution to follow under the class of flowing geometries constructed in \cite{Figueras:2012rb,Emparan:2013fha,Sun:2014xoa,Amado:2015uza,Megias:2015tva,Herzog:2016hob,Megias:2016vae,Sonner:2017jcf}.

\begin{figure}[h]
\centering
\includegraphics[width=0.4\linewidth]{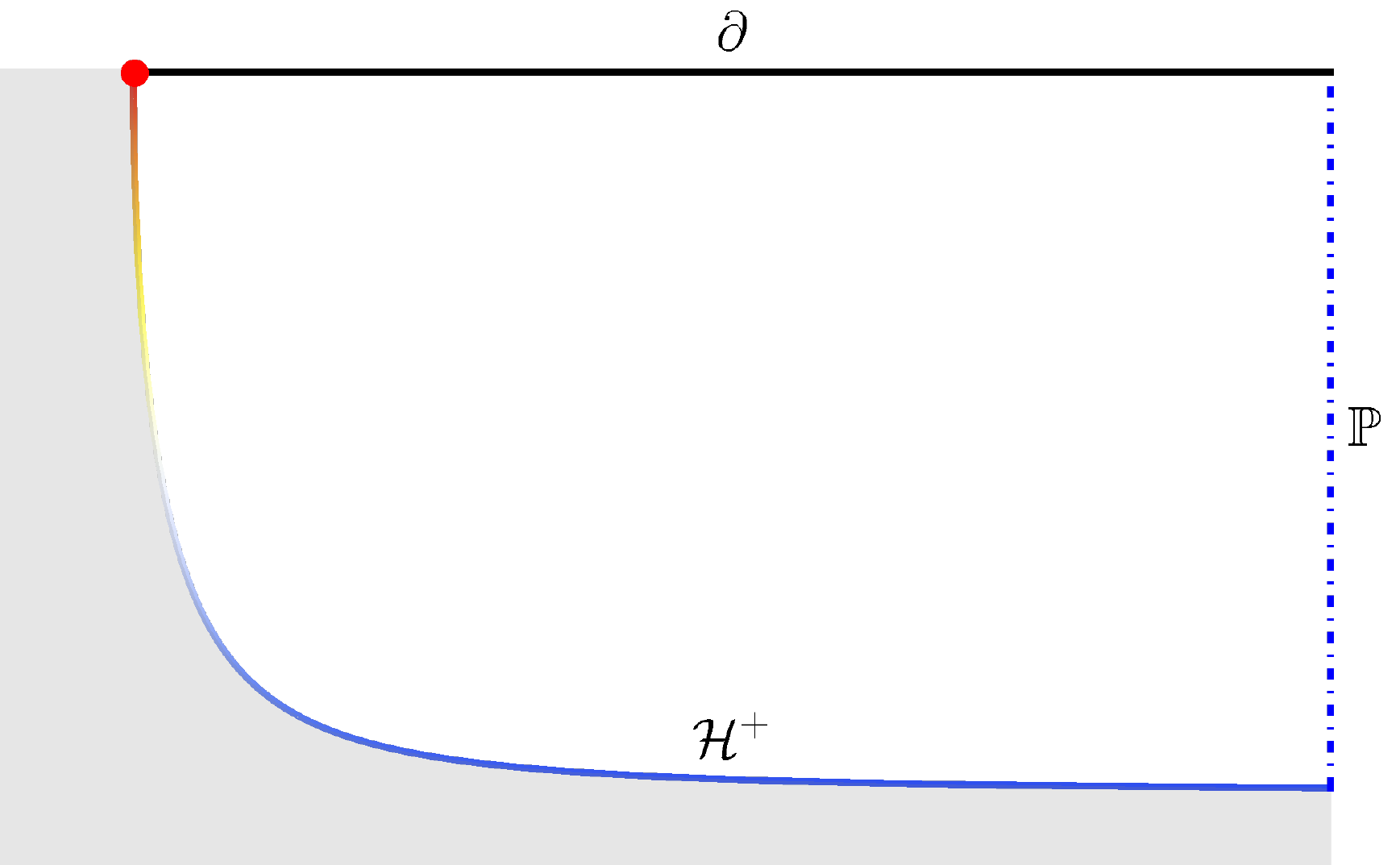}\hspace{1.5cm}
\includegraphics[width=0.4\linewidth]{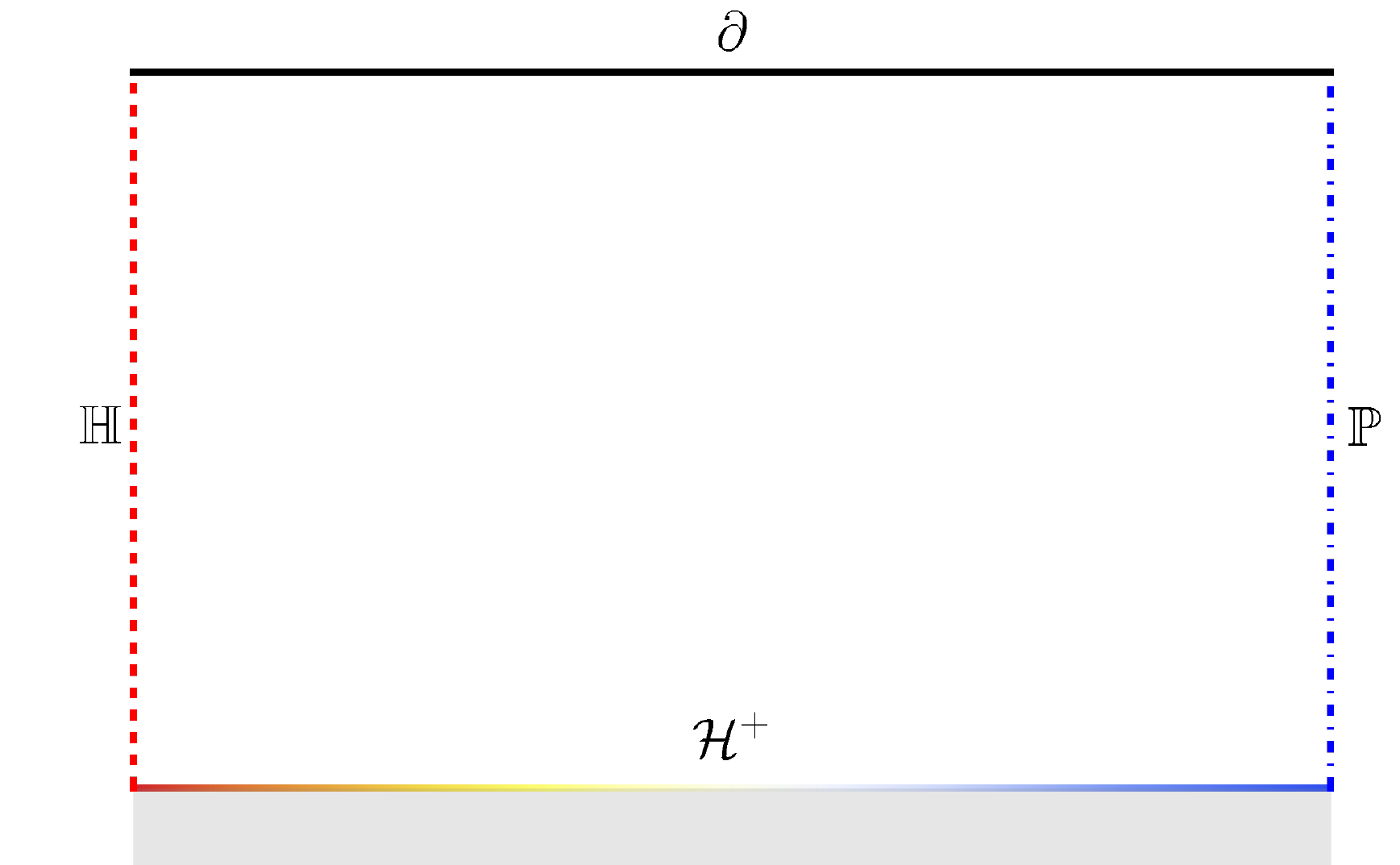}
\caption{An illustration of the coordinate transformation mapping the original black funnel triangular domain (left panel), into a square domain (right panel).}
\label{fig:blackfunnels}
\end{figure}

We begin by reviewing a coordinate system that is well adapted to the construction of black funnels \cite{Santos:2012he,Fischetti:2012vt}. In essence, a black funnel contains a single component horizon which connects the boundary black hole to an asymptotic region that is located infinitely far away from the boundary black hole  (see left panel of Fig.~\ref{fig:blackfunnels}). In this asymptotic region, the geometry should approach that of a standard five-dimensional planar black hole, which we denote on the left panel of Fig.~\ref{fig:blackfunnels} by $\mathbb{P}$. Thus, black funnels admit a natural triangular integration domain with three boundaries: a horizon, the planar black hole metric, and the conformal boundary.

Working with triangular integration domains can be tricky, so in \cite{Santos:2012he,Fischetti:2012vt} a new coordinate system was introduced, such that the point where the bulk future horizon $\mathcal{H}^+$ meets the boundary, was blown up into a line. By a careful inspection of the region where the bulk horizon meets the boundary horizon, one can show that the metric approaches that of a hyperbolic black hole. In fact this has to be the case for any horizon that meets the boundary, the reason being close to the conformal boundary the geometry is hyperbolic, up to subleading terms that are related to the holographic stress energy tensor. Moreover, the most general cohomogeneity one line element that is static and manifestly exhibits hyperbolic symmetry for each constant time hyperslice is that of a hyperbolic black hole. Each of these hyperbolic black holes is then determined by a choice of boundary black hole temperature, and bulk horizon temperature at the hyperbolic black hole horizon. We are interested in the situation where the boundary black hole has the same temperature as the bulk horizon temperature at the hyperbolic black hole horizon. The only such black hole is the zero energy hyperbolic black hole (which is isometric to pure AdS), and is given by
\begin{equation}
\mathrm{d}s_\mathbb{H}^2=\frac{L^2}{z^2}\left[-\left(1-z^2\right)\mathrm{d}\hat{t}^2+\frac{\mathrm{d}z^2}{1-z^2}+\mathrm{d}\eta^2+\sinh^2\eta\,\mathrm{d}\Omega_2^2\right]\,.
\label{eq:hyper}
\end{equation}
where here henceforth $\mathrm{d}\Omega_2^2$  will denote the unit radius round metric on $S^2$.

We want to find coordinates $(x,y)$ where at $x=0$ we approach the zero energy hyperbolic black hole with the same temperature as the boundary black hole $T_{\mathrm{BH}}$, at $x=1$ we want to impose that we approach a Schwarzschild black brane with some temperature $T_{\infty}$ and finally we want $y=0$ to denote the conformal boundary. Ideally the horizon would be an hypersurface with $y=1$, which was achieved in \cite{Fischetti:2012vt} by a careful choice of gauge. In this work we decided to follow more closely the approach devised in \cite{Figueras:2012rb}, but with some important differences which we will detail below.

Finally, at the conformal boundary we want to consider a geometry which is conformal to Schwarzschild, since this is the spacetime we want our field theory to live on. This means that close to $y=0$ the solution has to approach a Schwarschild black string
\begin{equation}
\mathrm{d}s^2=\frac{L^2}{y^2}\left[-\left(1-\frac{2M}{r}\right)\mathrm{d}t^2+\frac{\mathrm{d}r^2}{\displaystyle 1-\frac{2M}{r}}+r^2\mathrm{d}\Omega_2^2+\mathrm{d}y^2\right]\,,
\label{eq:stringads}
\end{equation}
with temperature
\begin{equation}
T_{\mathrm{BH}}=\frac{1}{8\pi M}=\frac{1}{4\pi r_0}\,,
\end{equation}
where we defined the Schwarzschild radius $r_0=2M$. Note that the temperature of the black string matches that of the Hawking temperature of a Schwarzschild black hole.

The general idea in both \cite{Fischetti:2012vt} and \cite{Figueras:2012rb} was to use the so called DeTurck method, which was first introduced in \cite{Headrick:2009pv}, and reviewed in detail in \cite{Wiseman:2011by,2016CQGra..33m3001D}. Recall we seek to construct solutions of the five-dimensional Einstein equation
\begin{equation}
R_{ab}+\frac{4}{L^2}g_{ab}=0\,,
\label{eq:einstein} 
\end{equation}
where $L$ is the AdS length scale. We will take latin indices to run over bulk spacetime dimensions, and greek indices to run over boundary spacetime dimensions. In order to apply the DeTurck method, we modify Eq.~(\ref{eq:einstein}) and consider instead
\begin{equation}
R_{ab}+\frac{4}{L^2}g_{ab}-\nabla_{(a}\xi_{b)}=0\,,
\label{eq:deturck}
\end{equation}
where $\xi^a\equiv g^{bc}[\Gamma^a_{bc}(g)-\Gamma^a_{bc}(\bar{g})]$, with $\bar{g}$ being a reference metric which should obey to the same Dirichlet boundary conditions as the physical metric $g$ we wish to find, $\Gamma^a_{bc}(\mathfrak{g})$ is the Christoffel connection associated with a metric $\mathfrak{g}$. Of course, solutions of Eq.~(\ref{eq:deturck}) will only coincide with solutions of Eq.~(\ref{eq:einstein}) if $\nabla_{(a}\xi_{b)}=0$. Under certain special circumstances one can show that solutions with $\xi^a\neq0$ cannot exist (see for instance \cite{Figueras:2011va,Figueras:2016nmo}), and as such solutions of Eq.~(\ref{eq:deturck}) will necessarily coincide with solutions of Eq.~(\ref{eq:einstein}). However, the solutions we seek do not satisfy the conditions of such theorems, so we need to check \emph{a posteriori} if $\xi$ approaches $0$ in the continuum limit.

There are many other circumstances where one cannot show that $\xi$ is necessarily zero on solutions of (\ref{eq:deturck}) (see for instance \cite{Donos:2014yya,Donos:2015eew,Costa:2017tug,Crisford:2017gsb,Horowitz:2019eum}). However, in many of these cases, one can show that the resulting system of equations is Elliptic, and as such one can trust local uniqueness of solutions to determine whether $\xi$ vanishes or not. The situation here is, however, more delicate. One can show that the system of equations we wish to solve does not appear Elliptic, instead the system appears to be of the mixed Elliptic-Hyperbolic type. This in turn means that we cannot use local uniqueness to distinguish between solutions with $\xi\neq0$ and solutions with $\xi=0$. We note however that this problem has an alternative formulation \cite{santos:2020} which does appear to have an Elliptic character.

We will now introduce our line element, and show that it reproduces all of the Dirichlet conditions detailed above. Our line element reads
\begin{multline}
\mathrm{d}s^2=\frac{L^2}{x\,y^2\,H(x)^2}\Bigg\{-x\,q_1\,\mathrm{d}v^2-2\,x\,q_2\,\mathrm{d}v\,\mathrm{d}y+q_5\,\mathrm{d}y^2+
\\
\frac{q_3}{4\,x(1-x)^4}\left[\mathrm{d}x+x\,(1-x)^2\,q_6\,\mathrm{d}y+x\,(1-x)^2\,q_7\,\mathrm{d}v\right]^2+\frac{q_4}{4\,(1-x)^2}\mathrm{d}\Omega_2^2\Bigg\}
\label{eq:crazy}
\end{multline}
For the reference metric, we will use the line element above with
\begin{align}
&q_1=(1-y^2)(1+x\,y^2)\nonumber
\\
&q_2=H(x)\nonumber
\\
&q_3=q_4=1\nonumber
\\
&q_i=0\quad\text{for}\quad i \in\{5,6,7\}.\nonumber
\end{align}
Finally, we will also choose $H(x)=1+\varpi\,x\,\sqrt{2-x^2}$. We shall see below that $\varpi$ will control the temperature of the black brane infinitely far way from the hole, and thus fix the temperature of the field theory reservoir.

We now discuss the thorny issue of boundary conditions. At $y=0$ we demand $q_1=q_3=q_4=1$, $q_2=H(x)$ and $q_i=0$ for $i\in\{5,6,7\}$. The metric then reads
\begin{equation}
\mathrm{d}s^2_{y=0}=\frac{L^2}{y^2H(x)^2}\Bigg[-x\,\mathrm{d}v^2-2\,x\,H(x)\,\mathrm{d}v\,\mathrm{d}y+\frac{\mathrm{d}x^2}{4\,x(1-x)^4}+\frac{\mathrm{d}\Omega_2^2}{4\,(1-x)^2}\Bigg]\,.
\end{equation}
We now take the following coordinate transformations
\begin{subequations}
\begin{align}
&v=\frac{t}{2\,r_0}-H(x)\,y
\label{eq:globaltime}
\\
& r = \frac{r_0}{1-x}
\\
& z=2\,r_0\,y\,H(x)
\end{align}
\end{subequations}
which reveals
\begin{equation}
\mathrm{d}s^2_{z=0}=\frac{L^2}{z^2}\left[-\left(1-\frac{r_0}{r}\right)\,\mathrm{d}t^2+\frac{\mathrm{d}r^2}{\displaystyle1-\frac{r_0}{r}}+r^2\mathrm{d}\Omega_2^2+\mathrm{d}z^2+\mathcal{O}(z)\right]\,.
\end{equation}
The line element above is that of a Schwarzschild black string in AdS (\ref{eq:stringads}). That is to say, the boundary metric is conformal to that of a Schwarzschild black hole. This is precisely what we want if the field theory is to live on a fixed Schwarzschild black hole background.

At $x=0$, we demand $q_1=1-y^2$, $q_2=q_3=q_4=1$ and $q_i=0$ for $i\in\{5,6,7\}$. This transforms (\ref{eq:crazy}) to
\begin{equation}
\mathrm{d}s^2_{x=0}=\frac{L^2}{y^2}\Big[-(1-y^2)\mathrm{d}v^2-2 \mathrm{d}v\,\mathrm{d}y+\frac{1}{4\,x^2}\mathrm{d}x^2+\frac{1}{4\,x}\mathrm{d}\Omega_2^2\Big]\,.
\label{eq:limitx0}
\end{equation}
We now define the following coordinate transformations
$$
x=e^{-2\,\eta}\,,\quad\mathrm{d}v=\mathrm{d}\hat{t}-\frac{\mathrm{d}y}{1-y^2}\quad \text{and}\quad y = z\,,
$$
which brings Eq.~(\ref{eq:limitx0}) to coincide with the large $\eta$ limit of Eq.~(\ref{eq:hyper}). We also recall that such black hole has temperature
\begin{equation}
T_{\mathbb{H}}=\frac{1}{4\,\pi\,r_0}=T_{\mathrm{BH}}\,,
\label{eq:tempH}
\end{equation}
measure by the time defined in Eq.~(\ref{eq:globaltime}).

Finally, we come to the more delicate boundary located at $x=1$. At this boundary we want the impose the metric of a planar Schwarzschild black brane with a different temperature than the line element (\ref{eq:limitx0}). At $x=1$, we demand $q_1=1-y^4$, $q_2=H(1)$ and $q_3=q_4=1$ and $q_i=0$ for $i\in\{5,6,7\}$, which brings the line element (\ref{eq:crazy}) to
\begin{equation}
\mathrm{d}s^2_{x=1}=\frac{L^2}{H(1)^2y^2}\Big[-(1-y^4)\mathrm{d}v^2-2 H(1)\mathrm{d}v\,\mathrm{d}y+\frac{1}{4\,(1-x)^4}\mathrm{d}x^2+\frac{1}{4\,(1-x)^2}\mathrm{d}\Omega_2^2\Big]\,.
\label{eq:limitx1}
\end{equation}
Now take
\begin{subequations}
\begin{align}
&x=1-\frac{1}{2H(1)r}
\\
& \mathrm{d}v=\mathrm{d}t-\frac{H(1)\mathrm{d}y}{1-y^4}
\end{align}
\end{subequations}
which transforms the metric given in Eq.~(\ref{eq:limitx1}) to
\begin{equation}
\mathrm{d}s^2_{x=1}=\frac{L^2}{y^2}\Big[-(1-y^4)\frac{\mathrm{d}t^2}{H(1)^2}+\frac{\mathrm{d}y^2}{1-y^4}+\mathrm{d}r^2+r^2\mathrm{d}\Omega_2^2\Big]\,.
\label{eq:planar}
\end{equation}
This metric describes a five-dimensional planar black hole with temperature
\begin{equation}
T_{\infty} = \frac{1}{2\,r_0H(1)\pi}\,,
\label{eq:tempP}
\end{equation}
measure by the time defined in Eq.~(\ref{eq:globaltime}).

Note that if we compare Eq.~(\ref{eq:tempH}) and Eq.~(\ref{eq:tempP}) we see that there is a gradient of temperatures between the two horizons unless, $H(1)=2$, \emph{i.e.} $\varpi = 1$. We will be interested in situations where $\varpi\neq1$, and in particular
\begin{equation}
\tau\equiv T_{\infty}/T_{\mathrm{BH}}=\frac{2}{1+\varpi}\neq1\,.
\end{equation}

One might wonder whether we are still missing a boundary condition in the interior. However, that is not the case. One can understand this in the following manner: the solution we are seeking is regular in ingoing coordinates, but badly singular in outgoing coordinates. This is enough to show that there are two possible solutions in the interior: one that blows up and one that does not. By working with a Chebyshev-Lobatto grid we automatically assume enough smoothness to guarantee that we only capture the smooth solution. The price we pay of not imposing a boundary condition in the interior is that we do not know \emph{a priori} where the horizon is. All we know is that it must join $y=1$ at both $x=0$ and $x=1$, due to the Dirichlet boundary conditions detailed above.
\section{\label{sec:hor}The horizon and its properties}
Horizons in general relativity are null-hypersurfaces, and can thus be written as the zero level sets of a function $\hat{h}$, which we choose to take the simple form $\hat{h}(x,y)\equiv y-P(x)$ \cite{Figueras:2012rb}
\begin{equation}
g^{ab}\nabla_a \hat{h}\nabla_b \hat{h}=0\,,
\label{eq:horizon}
\end{equation}
\emph{i.e.} $\mathrm{d}\hat{h}$ is null. Eq.~(\ref{eq:horizon}) yields and ODE for $P$, which one can readily solve. Furthermore, we know that $P(0)=P(1)=1$, either of which we can use as boundary conditions. Since the horizons we seek to construct are not Killing horizons, they will have unusual properties. For example, the expansion $\Theta$ and shear $\sigma$ of the horizon generators will be non-vanishing. Note, however, that since the horizon is a null-hypersurface, the rotation of the horizon generators can always be chosen to vanish even for non-Killing horizons.

In principle, computing $\Theta$ and $\sigma$ can be a relatively daunting task. However, in \cite{Fischetti:2012vt} a series of tricks where used to determine $\Theta$ and $\sigma$, as well as a choice of affine parameter $\lambda$ for the horizon generators. Since the properties of such quantities will play an important test of our numerics, we will review (and generalise slightly) the construction detailed in \cite{Fischetti:2012vt}.

The class of spacetimes we seek to find have a time translation symmetry and an $S^2$ symmetry. We shall use coordinates $\{v,x,y,\theta,\phi\}$, where $\partial/\partial v$ is the Killing vector field associated with time translations, $\phi$ and $\theta$ parametrise the $S^2$, and $y = P(x)$ on $\mathcal{H}^+$, the future event horizon. For the sake of notation, let us define the shorthand notation $\partial_I$, so that $\partial_v \equiv \partial/\partial v$, $\partial_\theta \equiv \partial/\partial \theta$ and $\partial_\phi \equiv \partial/\partial \phi$, and thus $I=v,\theta,\phi$.

$\mathcal{H}^+$ is $4$-dimensional, with a $3-$dimensional space of generators. We want to fix an $x$ coordinate on $\mathcal{H}^+$, and then uniquely label all horizon generators using the coordinates $(v,\theta,\phi)$ of their intersection with this surface. This should be possible provided that $\mathcal{H}^+$ is not Killing. We choose one horizon generator and pick an affine parameter $\lambda$. $\lambda$ can then be extended to a scalar function on $\mathcal{H}^+$ by requiring it to be independent of $v$, $\theta$ and $\phi$, so that $\lambda = \lambda(x)$. By symmetry, $\lambda$ will serve as an affine parameter for each geodesic.

Let $k^a$ be the tangent vector to the horizon generators with affine parameter $\lambda$. Since $\partial_I$ are all parallel to $\mathcal{H}^+$ we have $k \perp \partial_I$, for all $I$.

$\partial_I$ are deviations vectors for the geodesic congruence, since $k$ and $\partial_I$ commute. Therefore we have
\begin{equation}
k^c\nabla_c (\partial_I)^a=B^{a}_{\phantom{a}c}(\partial_I)^c
\end{equation}
where $B_{ab}=\nabla_{a}k_b$ is symmetric, since $k$ is hypersurface orthogonal.

Now define $h_{IJ}=\partial_I \cdot \partial_J$ for $I,J=v,\theta,\phi$. In the coordinates  $\{v,x,y,\theta,\phi\}$, we can identify $h_{IJ}$ with the $IJ$th component of the metric tensor through
\begin{equation}
h_{IJ}=g_{IJ}\,.
\end{equation}
The following is then true
\begin{align}
\frac{\mathrm{d}}{\mathrm{d}\lambda}h_{IJ} &=k^c\nabla_c(\partial_I \cdot \partial_J) \nonumber
\\
& = (\partial_J)_a B^{a}_{\phantom{a}c}(\partial_I)^c+(\partial_I)^a B_{a}^{\phantom{a}c}(\partial_J)_c \nonumber
\\
& = 2 B{ac} (\partial_J)^a(\partial_I)^c\nonumber
\\
& = 2B_{IJ}\,,
\end{align}
where $B_{IJ}$ denotes the $IJ$th component of $B$.

To obtain the expansion and shear of the congruence, it is necessary to work in the $3=5-2$ dimensional vector space $\hat{V}$ obtained by first restricting to vectors orthogonal to $k$, and then quotienting by an equivalence relation where two vectors are equivalent if they differ by a multiple of $k$. Tensors in the spacetime give rise to natural tensors in $\hat{V}$ if they obey the property that contracting any one index with $k_a$ or $k^a$ and the remainder with vectors or dual vectors having natural realisations in $\hat{V}$, gives zero. $B$, $g$ and $\partial_I$ all have this property. Furthermore, since $k$ and all $\partial_I$ are linearly independent, we have that all $\hat{\partial}_I$ are linearly independent on $\hat{V}$.

We can thus pick $\{\hat{\partial}_I\}$ as a basis for $\hat{V}$. For any tensor $T_{ab}$ naturally giving rise to a tensor in $\hat{V}$ , \emph{i.e.}
\begin{equation}
T_{ab} (\partial_I)^a (\partial_J)^b=\hat{T}_{ab} (\hat{\partial}_I)^a (\hat{\partial}_J)^b\,,
\end{equation}
we can identify $\hat{T}$ in $\hat{V}$ by reading off the components  $\{v,\theta,\phi\}$ of $T$. This is the case for the tensors $g$ and $B$.

The expansion and shear can now be determined using the usual formulae in terms of the quantities
\begin{subequations}
\begin{align}
&\hat{g}_{IJ}=g_{IJ}\,,
\\
&\hat{B}_{IJ}=B_{IJ}=\frac{1}{2}\frac{\mathrm{d}}{\mathrm{d}\lambda}g_{IJ}\,,
\end{align}
\end{subequations}%
where the final term means the derivative of the $IJ$th component of the metric along a geodesic, not the $IJ$the component of the covariant derivative of the metric.

It remains to find $\lambda$ as a function of $x$, but this can be achieved by solving Raychaudhuri’s equation written using the above expressions for $\hat{g}$ and $\hat{B}$ \cite{Fischetti:2012vt}. This gives an equation for $\lambda$ as a function of $x$ on $\mathcal{H}^+$:
\begin{subequations}
\begin{equation}
\frac{\mathrm{d}}{\mathrm{d}x}\log \left(\frac{\mathrm{d}}{\mathrm{d}x}\lambda\right)=\frac{1}{\left(h^{IJ}Dh_{IJ}\right)}\left[D\left(h^{IJ}Dh_{IJ}\right)+\frac{1}{2}h^{IK}h^{JL}\left(Dh_{IJ}\right)\left(Dh_{KL}\right)\right]\,,
\end{equation}
where
\begin{equation}
D\equiv\frac{\partial}{\partial x}+\frac{\mathrm{d}P}{\mathrm{d}x}\frac{\partial}{\partial y}\,.
\end{equation}
It is then possible to obtain $\hat{B}$ as
\begin{equation}
\hat{B}_{IJ}=\frac{1}{2}\frac{\mathrm{d}x}{\mathrm{d}\lambda}D g_{IJ}\,.
\end{equation}
\end{subequations}%
To compute the expansion and shear, we simply look at the trace and trace-free parts of $\hat{B}$, namely
\begin{equation}
\Theta = h^{IJ}\hat{B}_{IJ}\quad \text{and}\quad \sigma_{IJ}=\hat{B}_{IJ}-\frac{h_{IJ}}{3}\Theta\,.
\end{equation}
\section{\label{sec:holo}Extracting the Holographic-Stress energy tensor}
One of the most interesting quantities to extract from these solutions is the associated holographic stress energy tensor $\langle \mathrm{FP} |T^{\mu}_{\phantom{\mu}\nu}|\mathrm{FP}\rangle$, where $\mathrm{FP}$ stands for Frolov-Page states \cite{Frolov:1993fy}. We follow closely \cite{Balasubramanian:1998de,deHaro:2000vlm} whose starting point is to cast our solutions into Fefferman-Graham coordinates \cite{FG:1985,Graham:1999jg,Anderson:2004yi,Fefferman:2007rka}. Fortunately, we only need to perform this coordinate transformation asymptotically.

We start by determining the behaviour of all $q_{\tilde{I}}$, with $\tilde{I}\in\{1,\ldots,7\}$, close to $y=0$ by solving Eq.~(\ref{eq:deturck}) asymptotically. A careful analysis reveals the following asymptotic expansion
\begin{align}
q_{\tilde{I}}(x,y) = 1+\sum_{i=1}^5 q_{\tilde{I}}^{(i)}(x) y^i +\widetilde{q}_{\tilde{I}}(x) y^5 \log y+ \widehat{q}_{\tilde{I}}(x) y^{2+2\sqrt{3}}+o(y^{2+2\sqrt{3}})\,,
\label{eq:expandqI}
\end{align}
with $\{q_1^{(4)}(x),\widehat{q}_1(x),q_2^{(5)}(x),q_3^{(4)}(x),q_5^{(4)}(x),q_5^{(5)}(x),\widehat{q}_7(x)\}$ not being fixed by any local analysis of the Einstein-DeTurck equation, \emph{i.e.} these coefficients correspond to data that can only be determined once regularity deep in the bulk is imposed and the corresponding equations of motion are solved. The remaining coefficients are all determined as functions of $\{q_1^{(4)}(x),\widehat{q}_1(x),q_2^{(5)}(x),q_3^{(4)}(x),q_5^{(4)}(x),q_5^{(5)}(x),\widehat{q}_7(x),H(x)\}$ and their derivatives along $x$.

Imposing $\xi=0$ asymptotically imposes further constraints, which in turn imply the tracelessness and transversality of the holographic stress energy tensor. In addition, it imposes a linear relation between $\widehat{q}_1(x)$ and $\widehat{q}_7(x)$, which can be used to show that the terms proportional to $y^{2+2\sqrt{3}}$ are pure gauge. Note, however, that in the DeTurck gauge they are present, and change the expected convergence of the spectral collocation methods we used to solve the Einstein-DeTurck equation from exponential to power law.

To determine the holographic stress energy tensor, we define the following asymptotic change of coordinates
\begin{align}
&v=\frac{V-r_{\star}(w)}{2\,r_0}+\sum_{i=1}^5\alpha_i(w) z^i +\widetilde{\alpha}(x) z^5 \log z+ \widehat{\alpha}(w) z^{2+2\sqrt{3}}+o(z^{2+2\sqrt{3}})\,,\nonumber
\\
&x=w+\sum_{i=1}^5\beta_i(w) z^i +\widetilde{\beta}(x) z^5 \log z+ \widehat{\beta}(w) z^{2+2\sqrt{3}}+o(z^{2+2\sqrt{3}})\,,\label{eq:expandcoord}
\\
&y=\sum_{i=1}^5\gamma_i(w) z^i +\widetilde{\gamma}(x) z^5 \log z+ \widehat{\gamma}(w) z^{2+2\sqrt{3}}+o(z^{2+2\sqrt{3}})\,,\nonumber
\end{align}
where $V$ is the boundary Eddington-Finkelstein coordinate and $r_{\star}$ is the standard Schwarzschild tortoise coordinate defined as
\begin{equation}
r_{\star}(w) = \int_{w_1}^w\frac{r_0}{\tilde{w}(1-\tilde{w})^2} \mathrm{d}\tilde{w} \,,
\end{equation}
with $w_1\in(0,1)$ a constant. The coefficients $\alpha_i$, $\beta_i$, $\gamma_i$, $\widetilde{\alpha}$, $\widetilde{\beta}$, $\widetilde{\gamma}$, $\widehat{\alpha}$, $\widehat{\beta}$ and $\widehat{\gamma}$ are then determined by changing to Fefferman-Graham coordinates, where the five-dimensional line element takes a particularly simple form
\begin{equation}
\mathrm{d}s^2 = \frac{L^2}{z^2}\left[g_{\mu\nu}(x^\rho,z)\mathrm{d}x^\mu \mathrm{d}x^\nu+\mathrm{d}z^2\right]\,.
\end{equation}
The metric components $g_{\mu\nu}(x^\rho,z)$ themselves admit an expansion around $z=0$ of the form
\begin{equation}
g_{\mu\nu}(x^\rho,z)=\bar{g}_{\mu\nu}(x^\rho)+z^2\,A_{\mu\nu}(x^\rho)+z^4\, B_{\mu\nu}(x^\rho)+z^4\,\log z\, C_{\mu\nu}\,(x^\rho)+o(z^4)\,.
\end{equation}

We choose our conformal boundary metric, \emph{i.e.} $\bar{g}(x^\rho)$, to be Schwarzchild
\begin{equation}
\mathrm{d}s^2_{\partial}\equiv \bar{g}_{\mu\nu}(x^\rho)\mathrm{d}x^\mu \mathrm{d}x^\nu=-\left(1-\frac{r_0}{r}\right)\,\mathrm{d}V^2+2\mathrm{d}V\,\mathrm{d}r+r^2\mathrm{d}\Omega_2^2\,,\quad \text{and}\quad r=\frac{r_0}{1-w}\,.
\end{equation}
Since our choice of boundary metric is Ricci flat, one can show that $C=0$ \cite{deHaro:2000vlm}. Indeed, all the $\log$ terms in Eq.~(\ref{eq:expandqI}) and Eq.~(\ref{eq:expandcoord}) conspire to give a net $C=0$. Furthermore, as we have commented above, the terms proportional to $z^{2+2\sqrt{3}}$ in Eq.~(\ref{eq:expandcoord}) can be adjusted to show that the terms proportional to $ y^{2+2\sqrt{3}}$ in Eq.~(\ref{eq:expandqI}) are pure gauge, justifying why no such irrational powers appear in the Fefferman-Graham expansion. Following \cite{deHaro:2000vlm}, the holographic stress energy tensor can then be uniquely reconstructed from $\bar{g}$, $A$ and $B$ \cite{deHaro:2000vlm}
\begin{equation}
\langle \mathrm{FP} |T_{\rho\epsilon}|\mathrm{FP}\rangle=\frac{L^3}{4\pi G_5}\Bigg[B_{\rho\epsilon}+\frac{1}{8}\left(A_{\mu\nu}A^{\mu\nu}-A^2\right)\bar{g}_{\rho\epsilon}-\frac{1}{2}A_{\rho\mu}A_{\epsilon}^{\phantom{\epsilon}\mu}+\frac{1}{4}A_{\rho\epsilon}A\Bigg]\,,
\end{equation}
where $A\equiv \bar{g}_{\mu\nu}A^{\mu\nu}$. Since the boundary metric is Ricci flat, there are no ambiguities in defining $\langle \mathrm{FP} |T_{\rho\epsilon}|\mathrm{FP}\rangle$. Finally, following the standard AdS/CFT dictionary \cite{Maldacena:1997re} we identify
\begin{equation}
G_5=\frac{\pi}{2}\frac{L^3}{N^2}\,.
\end{equation}

One quantity of interest to compute, once the holographic stress energy tensor is determined, is the heat flux $\Phi$. This quantity is interpreted here as the energy flux integrated over a two-sphere of constant $r$, that is to say
\begin{equation}
\Phi = -\int_{S^2_r}\xi^\mu \langle \mathrm{FP} |T_{\mu\nu}|\mathrm{FP}\rangle n^\nu \sqrt{-h}\,\,\mathrm{d}^2x\,,
\label{eq:flux}
\end{equation}
where $\xi^{\mu} =(\partial/\partial V)^{\mu}$, $\sqrt{-h}$ is the volume element induced on a constant $r$ slice and $n^{\nu}$ an outward unit normal to $S^2_r$. By virtue of the conservation of the stress energy tensor and since $\xi^{\mu}$ is a Killing vector, $\Phi$ is independent of $r$. With our conventions, an outgoing flux of radiation corresponds to $\Phi>0$ and an ingoing flux to $\Phi<0$. We expect $\Phi$ to be positive if $T_{\infty}<T_{\mathrm{BH}}$, \emph{i.e.} $\varpi>1$, to vanish for the Hartle-Hawking state $\varpi=1$, and to be negative for $\varpi<1$. In the subsequent section we will confirm this behaviour numerically.

\section{\label{sec:res}Results}
We start by studying the geometry of the future horizon $\mathcal{H}^+$. To help us gain intuition, we construct isometric embeddings of spatial cross sections of $\mathcal{H}^+$ into hyperbolic space. These are specially useful if we want to visualize where the horizon bulges out. Since spatial cross-sections of $\mathcal{H}^+$ are three-dimensional, we seek to construct embeddings into $\mathbb{H}_4$. We foliate four-dimensional hyperbolic space using three-dimensional flat space (\emph{i.e.} Poincar\'e slicing of Euclidean AdS)
\begin{equation}
\mathrm{d}s^2_{\mathbb{H}_4}=\frac{\tilde{L}^2}{Z^2}\left(\mathrm{d}Z^2+\mathrm{d}R^2+R^2 \mathrm{d}\Omega^2_2\right)\,.
\label{eq:hyper4}
\end{equation}
One then searches for an embedding of the form $(Z(x),R(x))$, that is to say the pull-back of Eq.~(\ref{eq:hyper4}) to a parametrised surface $(Z(x),R(x))$, which gives the following induced metric
\begin{equation}
\mathrm{d}\hat{s}^2_{\mathbb{H}_4}=\frac{\tilde{L}^2}{Z(x)^2}\left\{\left[Z^\prime(x)^2+R^\prime(x)^2\right]\mathrm{d}x^2+R(x)^2 \mathrm{d}\Omega^2_2\right\}\,.
\label{eq:hyper4R}
\end{equation}
We can now compare this line element with the pull back of the metric for the flowing funnel, \emph{i.e.}  Eq.~(\ref{eq:crazy}), induced on the intersection of the the future event horizon $\mathcal{H}^+$, identified via solving Eq.~(\ref{eq:horizon}), with a partial Cauchy surface of constant $v$ and read off a nonlinear first order equation for $Z(x)$. We fix the boundary conditions by demanding $Z(1)=1$. In making the identification between line elements we also set $\tilde{L}=L$. The advantage of this embedding is that a black string, see Eq.~(\ref{eq:stringads}), appears as a line of constant $R$, whereas a five-dimensional planar black hole, see Eq.~(\ref{eq:planar}), appears as a line of constant $Z$. Thus, a black funnel should naturally interpolate between these two curves.

In addition, we shall also be interested in constructing isometric embeddings of the ergosurfaces of our solutions. These surfaces are defined as surfaces for which the norm $\|\partial/\partial v\|^2$ becomes spacelike. Inside the ergosurface, that is to say in the \emph{ergoregion}, the character of Eq.~(\ref{eq:deturck}) changes from elliptic to hyperbolic. This is the reason why the existence of these surfaces is at the core of the difficulties in trying to prove that our numerical method ensures the absence of DeTurck solitons. We shall follow the same strategy and also embed the ergosurfaces into four-dimensional hyperbolic space.

\begin{figure}[h]
\centering
\includegraphics[width=\linewidth]{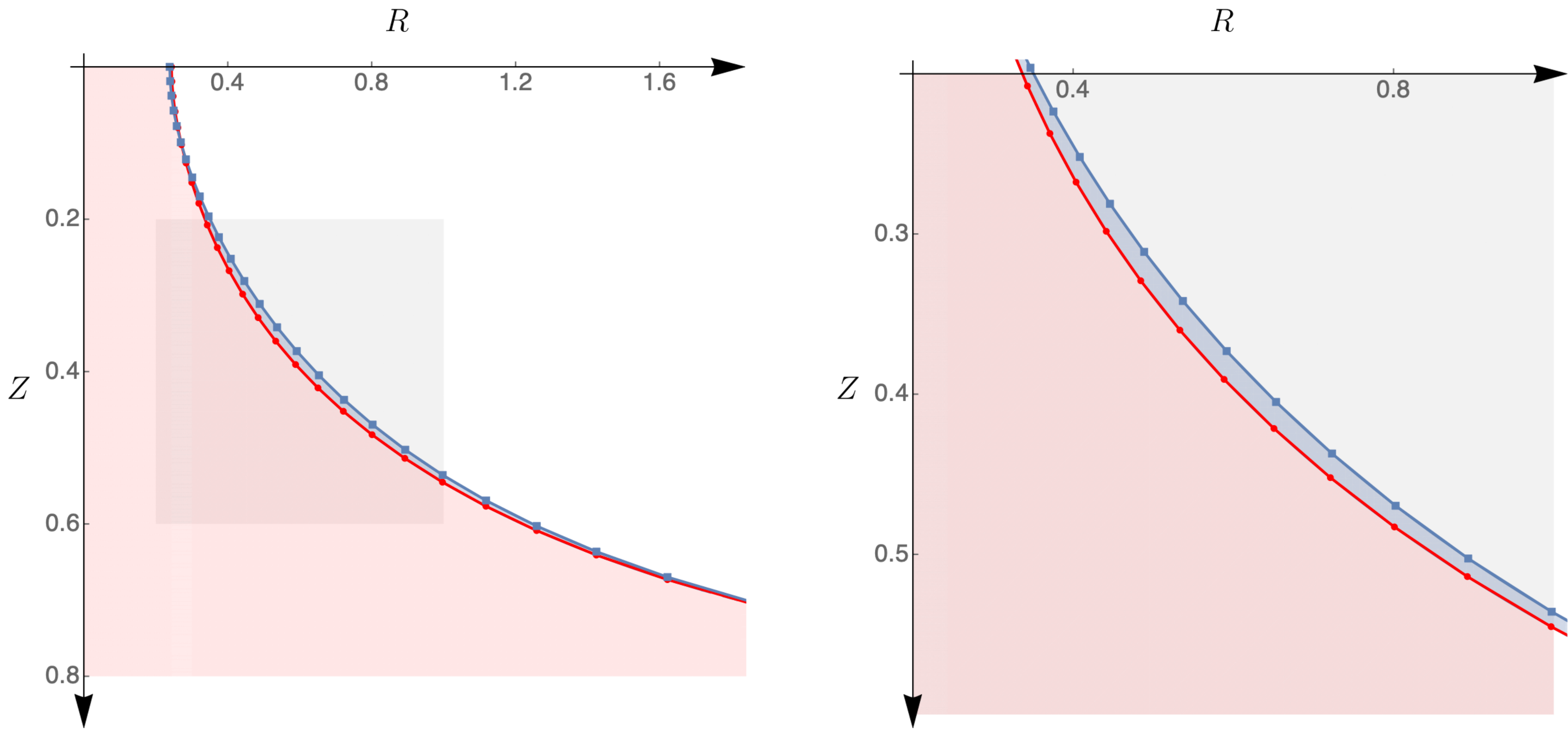}
\caption{Isometric embeddings of the spatial cross section of $\mathcal{H}^+$ (red disks) and the ergosurface (blue squares). These curves were generated with $T_{\infty}/T_{\mathrm{BH}}= 2.5$. The right panel shows a zoom of the shaded region on the left, so that one can more easily identify the ergoregion. The red shaded region corresponds to the interior of the black funnel, and the blue shaded region to the ergoregion.}
\label{fig:embeddings}
\end{figure}

In Fig.~\ref{fig:embeddings} we can observe the embedding of a partial Cauchy surface of constant $v$ with $\mathcal{H}^+$ (blue disks) and of the ergosurface (orange squares). The ergoregion remains very narrow even when $T_{\infty}/T_{\mathrm{BH}}= 2.5$. The same is true in the opposite limit, \emph{i.e.} $T_{\infty}/T_{\mathrm{BH}}\sim 0.6$. As expected, the ergoregion shrinks down to zero size at the boundary ($y=0$) and at the planar black hole asymptotic region ($x=1$). The existence of the bulk ergoregion also raises questions about the stability of the solutions we have just found, specially in light of \cite{Green:2015kur}. However, we note that in \cite{Green:2015kur} it was essential that the spatial cross section of the boundary metric was (metrically) a round sphere. A comprehensive analysis of the linear-mode stability of our solutions is outside the scope of this manuscript, though we plan to return to it in the near future.

Perhaps the most important quantity to extract is the stress energy tensor and its associated Hawking flux $\Phi$ defined in Eq.~(\ref{eq:flux}). Within our symmetry class, the holographic stress energy tensor has four non-zero independent components. In addition, it should be traceless and covariantly conserved, which gives three constraints amongst these four components. Thus, the full stress energy tensor is determined by, say, $\langle T_{VV}\rangle \equiv \langle \mathrm{FP} |T_{VV}|\mathrm{FP}\rangle$. Using conservation of the holographic stress energy tensor, it is also simple to show that if $\langle \mathrm{FP} |T_{Vw}|\mathrm{FP}\rangle\neq0$ on $\mathcal{H}^+$, and the holographic stress energy tensor is smooth on $\mathcal{H}^+$, then $\left.\langle T_{VV}\rangle\right|_{\mathcal{H}^+}\neq 0$. This is what we expect to happen when $T_{\infty}/T_{\mathrm{BH}}\neq1$.

\begin{figure}[h]
\centering
\includegraphics[width=\linewidth]{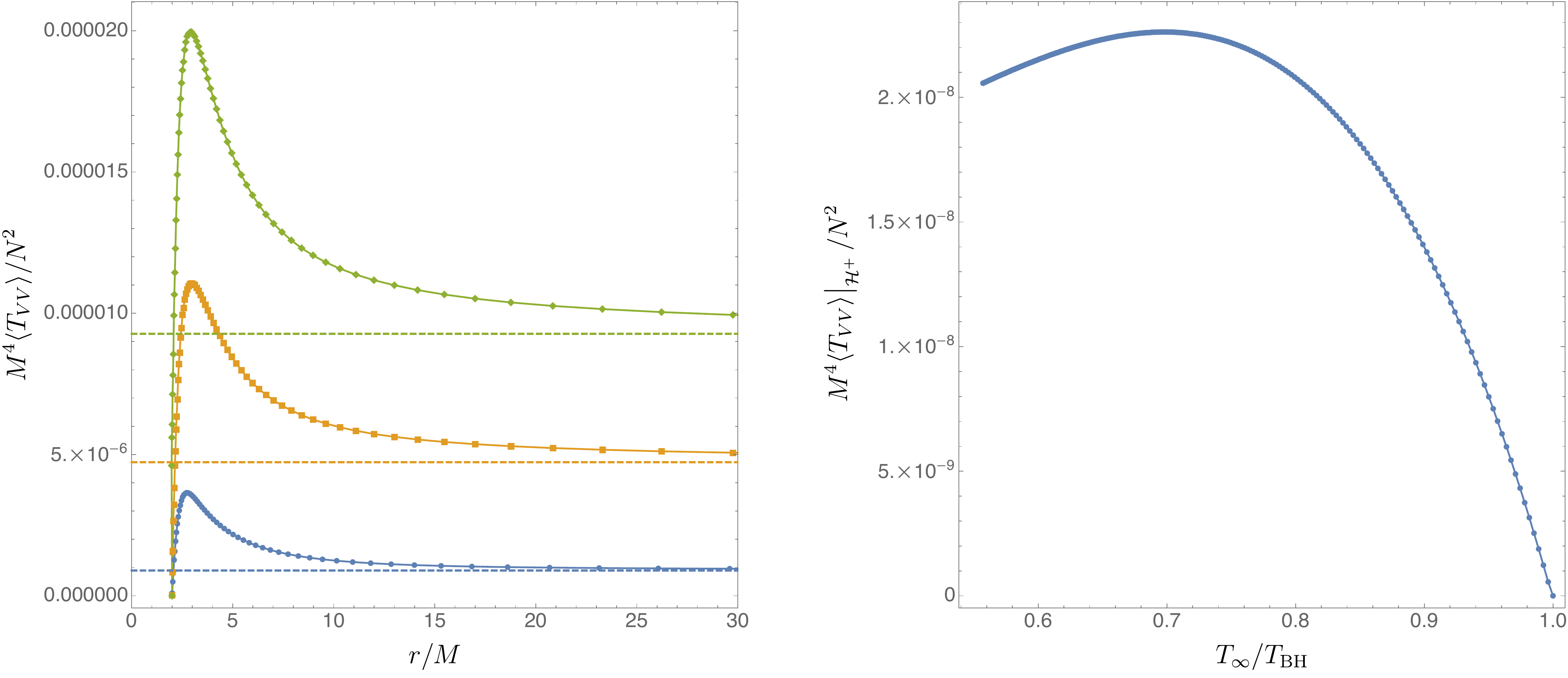}
\caption{{\bf Left: }The null-null component of the holographic stress energy tensor, $\langle T_{VV}\rangle$, as a function of $r/M$. Each curve corresponds to a different value of $T_{\infty}/T_{\mathrm{BH}}$. From bottom to top, we have $T_{\infty}/T_{\mathrm{BH}}=0.557103\,, 0.844773\,, 1$. The horizontal dashed lines indicate the asymptotic value of $\langle T_{VV}\rangle$ computed for a planar black hole at temperature $T_{\infty}$. {\bf Right:} The null-null component of the holographic stress energy tensor evaluated on $\mathcal{H}^+$, $\left.\langle T_{VV}\rangle\right|_{\mathcal{H}^+}$, as a function of several values of $T_{\infty}/T_{\mathrm{BH}}$. As expected, $\left.\langle T_{VV}\rangle\right|_{\mathcal{H}^+}$ is non-zero except for the Hartle-Hawking state corresponding to $T_{\infty}/T_{\mathrm{BH}}=1$. In both plots we have used $r_0=2M$.}
\label{fig:stress}
\end{figure}

In Fig.~\ref{fig:stress} we plot $\langle T_{VV}\rangle$ for several values of $T_{\infty}/T_{\mathrm{BH}}$. On the left panel of Fig.~\ref{fig:stress} blue disks, orange squares and green diamonds correspond to $T_{\infty}/T_{\mathrm{BH}}=0.557103\,, 0.844773\,, 1$, respectively. The right panel of Fig.~\ref{fig:stress} shows $\left.\langle T_{VV}\rangle\right|_{\mathcal{H}^+}$ and, as anticipated, it is non-zero except when $T_{\infty}/T_{\mathrm{BH}}=1$ (the Hartle-Hawking state). For the Hartle-Hawking state, we recover the results of \cite{Santos:2012he,Fischetti:2016oyo}.

\begin{figure}[h]
\centering
\includegraphics[width=\linewidth]{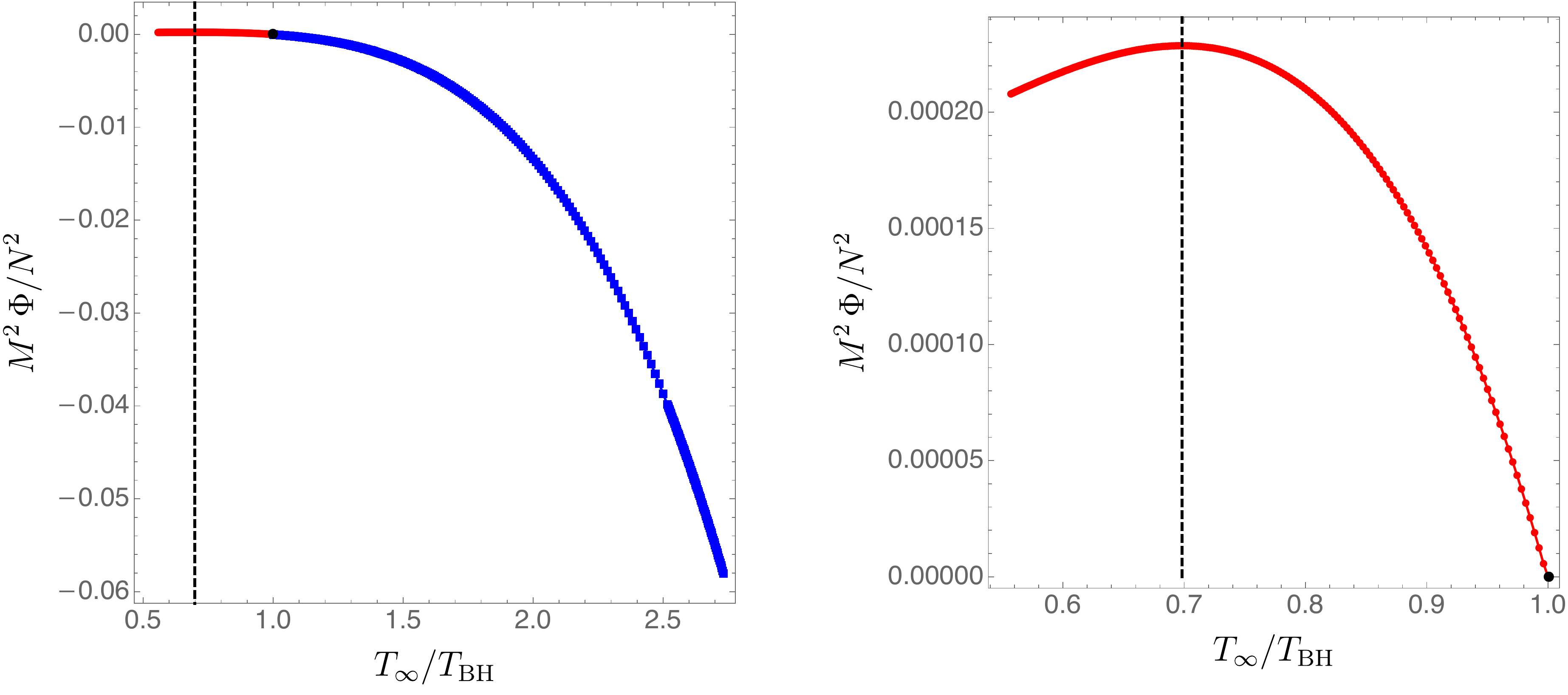}
\caption{Hawking flux $\Phi$ plotted as a function of $T_{\infty}/T_{\mathrm{BH}}$. The right panel shows a zoom of the left panel in the region $T_{\infty}/T_{\mathrm{BH}}\leq1$. $\Phi$ does not appear to behave monotonically in $T_{\infty}/T_{\mathrm{BH}}$, and in particular it has a maximum at $T_{\infty}/T_{\mathrm{BH}}\equiv \tau_{\max}\approx 0.698282$, which is marked as a vertical dashed line in both left and right panels. The red disks indicate when $\Phi>0$, and the blue squares when $\Phi<0$. The Hartle-Hawking state is represented by a black disk.}
\label{fig:hawkingflux}
\end{figure}

We now turn our attention to the Hawking flux $\Phi$ defined in Eq.~(\ref{eq:flux}), which we plot in Fig.~\ref{fig:hawkingflux}. The right panel shows a zoom of the left panel in the region $T_{\infty}/T_{\mathrm{BH}}\leq1$. As expected $\Phi<0$ for $T_{\infty}/T_{\mathrm{BH}}>1$, meaning that there is an influx of radiation into the black hole. The temperature of the heat bath provides an energy reservoir that sources the geometry. For the Hartle-Hawking state, $T_{\infty}/T_{\mathrm{BH}}=1$, there is no net flux of Hawking particles, so that $\Phi=0$. This is marked as a black disk in Fig.~\ref{fig:hawkingflux}. Finally, we see a somehow surprising result when $T_{\infty}/T_{\mathrm{BH}}<1$. This is indeed when we expect the black hole to radiate Hawking quanta, and as such $\Phi>0$. However, $\Phi$ does not seem to be monotonic in $T_{\infty}/T_{\mathrm{BH}}$. Instead, it initially grows with decreasing $T_{\infty}/T_{\mathrm{BH}}$, but it reaches a maximum at $T_{\infty}/T_{\mathrm{BH}}\equiv \tau_{\max}\approx 0.698282$. This value is marked in Fig.~\ref{fig:hawkingflux} as a vertical dashed black line. At the moment, we have no field theory understanding for this non-monotonic behaviour of $\Phi$.
Using the results of \cite{Figueras:2011he}, we have some preliminary results indicating that the flowing geometry is Gregory-Laflamme unstable \cite{Gregory:1993vy} in the region $T_{\infty}/T_{\mathrm{BH}}<\tau_{\max}$. It is natural to expect that the endpoint of such instability is one of the black droplets found in \cite{Santos:2014yja}. Such evolution will necessarily involve a violation of the weak cosmic censorship conjecture \cite{Penrose:1969pc}, alike the non-linear evolution of the Gregory-Laflamme instability \cite{Lehner:2010pn}. This is not the first time that violations of weak cosmic censorship conjecture play a role in the AdS context, see for instance \cite{Niehoff:2015oga,Dias:2016eto,Horowitz:2016ezu,Crisford:2017zpi,Marolf:2019wkz}, although it seems one of the easiest scenarios to actually sort out the evolution numerically using the methods outlined in \cite{Chesler:2013lia,Balasubramanian:2013yqa}.

\begin{figure}[h]
\centering
\includegraphics[width=\linewidth]{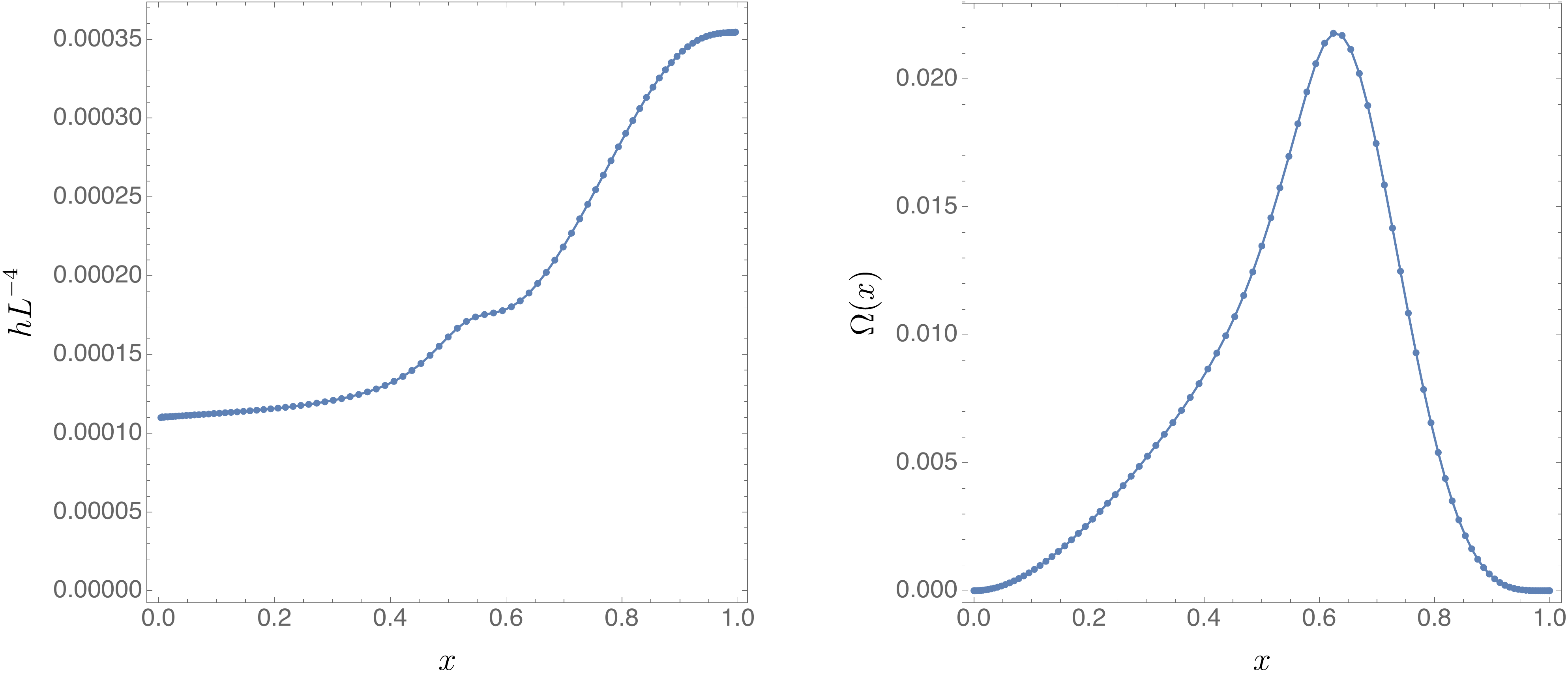}
\caption{{\bf Left:} The determinant $h$ as a function of $x$ on $\mathcal{H}^+$ plotted for fixed $T_{\infty}/T_{\mathrm{BH}}\approx0.557103$. {\bf Right:} The coordinate velocity $\Omega(x)$ on $\mathcal{H}^+$ plotted as a function of $x$ for fixed $T_{\infty}/T_{\mathrm{BH}}\approx0.557103$.}
\label{fig:det}
\end{figure}

We now discuss the properties of this novel class of horizons. For simplicity, we restrict our discussion to the region of moduli space $T_{\infty}/T_{\mathrm{BH}}<1$. We begin by studying the behaviour of $h_{IJ}$ itself, and in particular its determinant $h\equiv \mathrm{det} h_{IJ}$. We plot this quantity on the left panel of Fig.~\ref{fig:det} for $T_{\infty}/T_{\mathrm{BH}}\approx0.557103$. $h$ is clearly a monotonically increasing function of $x$, showing that $x$ increases towards the future along the future horizon, and thus that the past horizon lies at $x=0$. We note that in this region the temperature of the black hole is higher than that of the heat reservoir.

To understand the direction of the flow, we can also determine the horizon coordinate velocity $\Omega(x)$ along $\mathcal{H}^+$. To do so, we can look at the pull-back of our line element (\ref{eq:crazy}) to the future horizon $\mathcal{H}^+$ (see Eq.~(\ref{eq:horizon}))
\begin{align}
\mathrm{d}s^2_{\mathcal{H}^+}& =\frac{L^2}{x\,P^2\,H^2}\Bigg\{-x\,\tilde{q}_1\,\mathrm{d}v^2-2\,x\,\tilde{q}_2\,P^\prime\mathrm{d}v\,\mathrm{d}x+\tilde{q}_5\,{P^\prime}^2\mathrm{d}x^2+\nonumber
\\
& \quad\quad \frac{\tilde{q}_3}{4\,x(1-x)^4}\left[\mathrm{d}x+x\,(1-x)^2\,\tilde{q}_6\,P^\prime\mathrm{d}x+x\,(1-x)^2\,\tilde{q}_7\,\mathrm{d}v\right]^2+\frac{\tilde{q}_4}{4\,(1-x)^2}\mathrm{d}\Omega_2^2\Bigg\} \nonumber
\\
& = \frac{L^2}{4\,x\,P^2\,H^2}\Bigg\{\left[\frac{\tilde{q}_3 \left(1+(1-x)^2 x\,\tilde{q}_5 P^\prime\right)^2}{(1-x)^4 x}+4\,x\,\tilde{q}_7 {P^\prime}^2\right](\mathrm{d}x-\Omega(x) \mathrm{d}v)^2+\frac{\tilde{q}_4}{(1-x)^2}\mathrm{d}\Omega_2^2\Bigg\}\,.
\label{eq:pullback}
\end{align}
where the last equality follows from the fact that $\mathcal{H}^+$ is null, $\tilde{q}_{\tilde{I}}=q_{\tilde{I}}(x,P(x))$ and $^\prime$ denotes differentiation with respect to $x$. The last equality in Eq.~(\ref{eq:pullback}) implicitly defines the coordinate velocity $\Omega(x)$. We have plotted $\Omega(x)$ for $T_{\infty}/T_{\mathrm{BH}}\approx0.557103$ on the right panel of Fig.~(\ref{fig:det}), and we find $\Omega(x)\geq0$. Indeed, we find that $\Omega(x)\geq0$ so long as $T_{\infty}/T_{\mathrm{BH}}\leq1$, and negative otherwise. Both the sign of $\Omega$ and the monotonicity properties of $h$, indicate that the past horizon is located at the black hole horizon (the hotter reservoir). This is to contrast with the coordinate choice made in \cite{Fischetti:2012ps,Fischetti:2012vt}, in which the cooler horizon appears to be closer to the past horizon $\mathcal{H}^-$.

\begin{figure}[h]
\centering
\includegraphics[width=\linewidth]{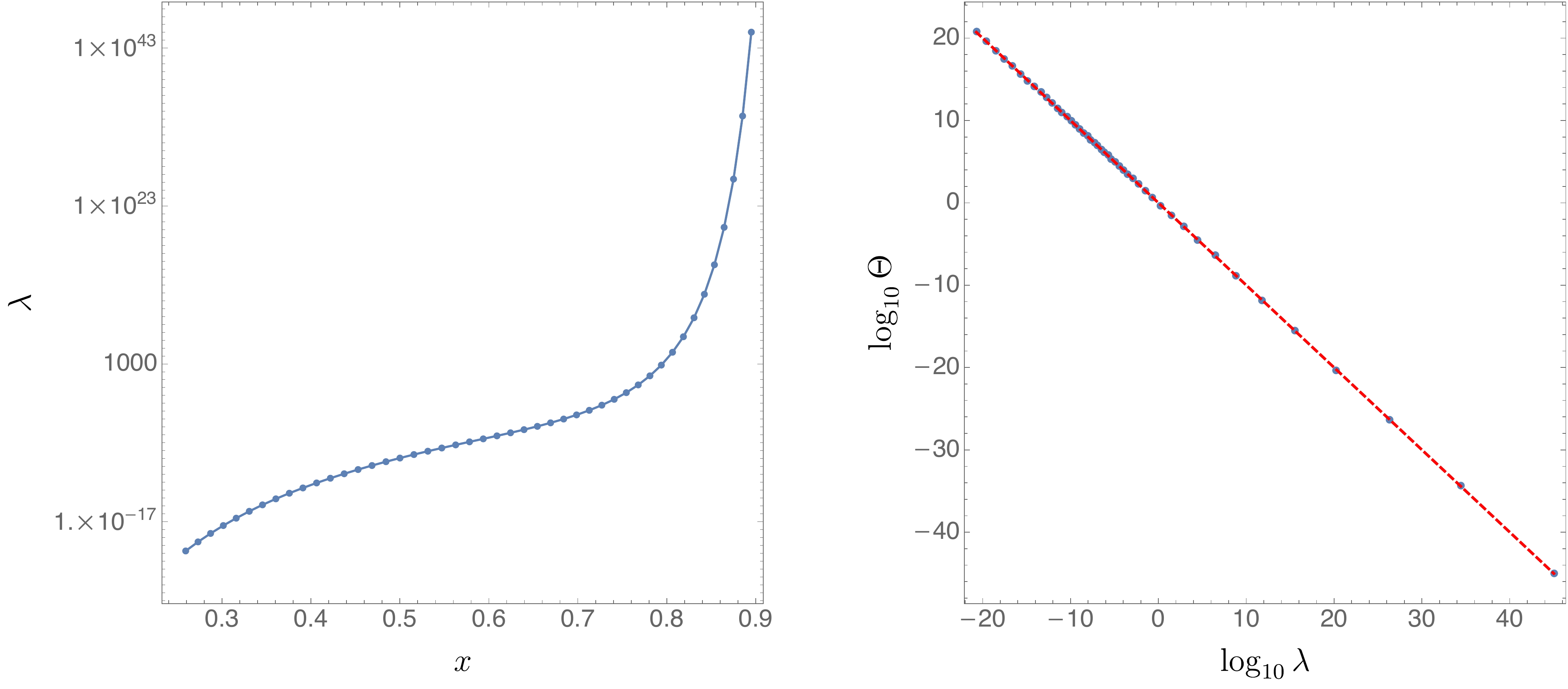}
\caption{{\bf Left:} Logarithmic plot of the affine parameter $\lambda$ as a function of $x$ on $\mathcal{H}^+$. {\bf Right:} $\log_{10}-\log_{10}$ plot of the expansion $\Theta$ as a function of $\lambda$. The dashed red line indicates a best fit consistent with $ \log_{10}\Theta=0.00195019 -  \log_{10}\lambda$. Both panels were generated with $T_{\infty}/T_{\mathrm{BH}}\approx0.729262$.}
\label{fig:expansion}
\end{figure}

Next, we look at the behaviour of the horizon expansion $\Theta$ as a function of $\lambda$. According to Raychaudhuri's equation and the definition of $\Theta$, it better be that $\Theta>0$, $\mathrm{d}\Theta/\mathrm{d}\lambda<0$ and that $\Theta$ approaches zero for large $\lambda$. On the left panel of Fig.~\ref{fig:expansion} we plot the affine parameter $\lambda$ as a function of $x$ on the future event horizon. Recall that $x$ is one of the coordinates introduced in Eq.~(\ref{eq:crazy}). As expected, $\lambda$ is a monotonically increasing function of $x$, with $x=0$ locating the past horizon. On the right panel, we show the expansion $\Theta$ as a function of the affine parameter $\lambda$, and we confirm that $\mathrm{d}\Theta/\mathrm{d}\lambda<0$. In order to show that this is the case using Raychaudhuri's equation, one needs to use the equations of motion. We thus see this as the most solid test of our numerical procedures. 
On the right panel of Fig.~\ref{fig:expansion} we also show, as a dashed red line, a linear fit in the $(\log_{10}\lambda,\log_{10}\Theta)$ plane, which yields $ \log_{10}\Theta=0.00195019 -  \log_{10}\lambda$. The fit seems to work for all ranges of $x$, so there is a clear indication that $\Theta$ is diverging as $\lambda^{-1}$ at the past horizon (located at $\lambda=0$), signalling the presence of a caustic. This caustic, in turn, gives rise to a curvature singularity. As noted in \cite{Fischetti:2012vt}, the easiest way to see this is to note that for any Killing field, the Ricci identify for vectors implies that
\begin{equation}
\nabla_a \nabla_b K^c = R^{c}_{\phantom{c}bad}K^d\,.
\label{eq:ricci}
\end{equation}
In particular, this will be true for any Killing vector on the two-sphere, say $K = \partial/\partial \phi$. It is easy to see that for any such Killing vector $\| K\|^2$ diverges at $x=0$, \emph{i.e.} when $\lambda=0$. But since $K$ obeys the second order partial differential equation (\ref{eq:ricci}), it can only diverge at finite affine parameter $\lambda=0$ if $R_{abcd}$ diverges at $\lambda=0$ in all orthonormal frames.

\begin{figure}[h]
\centering
\includegraphics[width=.57\linewidth]{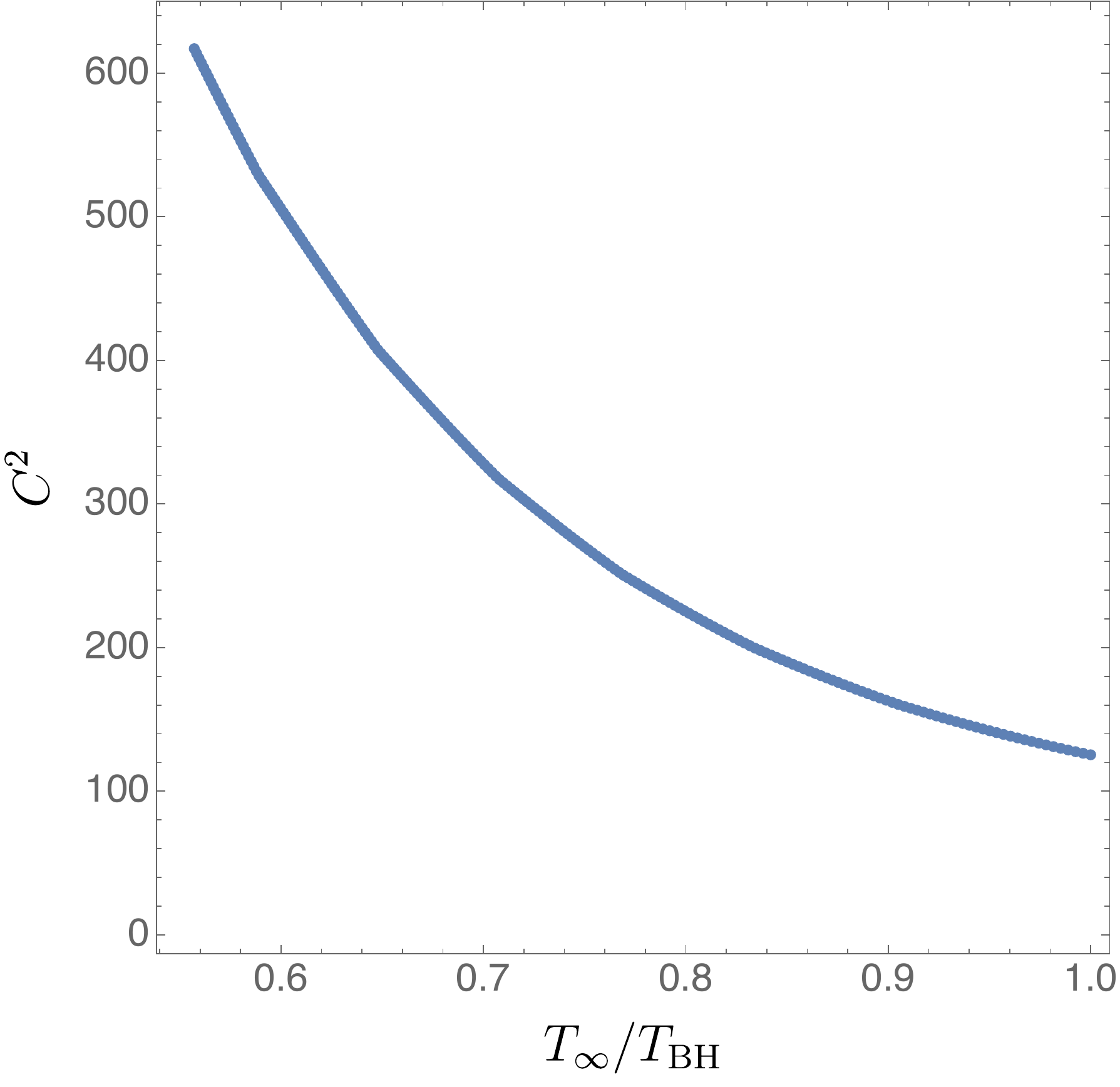}
\caption{$C^2\equiv \max_{\mathcal{C}}\left|C_{abcd}C^{abcd}\right|$, where $\mathcal{C}$ stands for domain of outer communications, as a function of $T_{\infty}/T_{\mathrm{BH}}$. We see no apparent divergent behaviour as we lower the temperature ratio $T_{\infty}/T_{\mathrm{BH}}$ .}
\label{fig:weyl}
\end{figure}

Finally, we end with a comment regarding the behaviour of our solutions close to the minimal temperature ratio $T_{\infty}/T_{\mathrm{BH}}=0.557103$ that we have reached. In Fig.~\ref{fig:weyl} we monitor the behaviour of $C^2\equiv \max_{\mathcal{C}}\left|C_{abcd}C^{abcd}\right|$, as a function of $T_{\infty}/T_{\mathrm{BH}}$, where $\mathcal{C}$ stands for domain of outer communications and $C$ is the five-dimensional Weyl tensor. Throughout moduli space, we find no evidence for a divergent behaviour. We suspect the reason why we cannot reach lower temperatures is purely numerical, and is related to the existence of large gradients inside the future event horizon.
\section{\label{sec:dis}Discussion}
We constructed the holographic dual of the Frolov-Page states \cite{Frolov:1993fy} for $\mathcal{N}=4$ SYM with gauge group $SU(N)$ and asymptotic temperature $T_{\infty}$, at large 't Hooft coupling and infinite $N$ on a Schwarzschild black hole background with temperature $T_{\mathrm{BH}}$. When $T_{\mathrm{BH}}=T_{\infty}$, the Frolov-Page states reduce to the Hartle-Hawking state, and when $T_{\infty}\to0$ one recovers the Unruh state. The resulting bulk geometries are of the flowing type, and thus similar in nature to the ones first uncovered in \cite{Fischetti:2012vt,Figueras:2012rb}. They possess spherical symmetry and a stationary Killing field $\partial/\partial v$ as well as an horizon $\mathcal{H}^+$ whose spatial cross sections are non-compact. However, the horizon is not a Killing horizon and in particular it is not generated by any linear combination of Killing fields. This is not in contradiction with the standard rigidity theorems \cite{Hawking:1971vc,Hawking:1973uf,Hollands:2006rj}, since the latter assume horizons with compactly generated spatial cross sections.

When $T_{\mathrm{BH}}\neq T_{\infty}$ we find a net flux of Hawking radition, being outgoing whenever $T_{\mathrm{BH}}>T_{\infty}$, and ingoing otherwise. For the Hartle-Hawking state, corresponding to $T_{\mathrm{BH}}=T_{\infty}$, the flux vanishes identically, as expected, and the results of \cite{Santos:2012he} are recovered. These novel flowing horizons have unfamiliar properties, such as a non-vanishing expansion $\Theta$. We have studied how $\Theta$ varies as a function of $T_{\mathrm{BH}}/T_{\infty}$, and find it is positive and extends to the future along each null generator. The horizon generators extend to infinite affine parameter $\lambda$ in the far future, but reach a caustic (at finite affine parameter $\lambda$. We have computed $C_{abcd}C^{abcd}$ and it seems to remain bounded in the domain of outer communications, indicating that the caustic located at $\lambda=0$ is likely to be tidal singularity.

\begin{figure}[h]
\centering
\includegraphics[width=\linewidth]{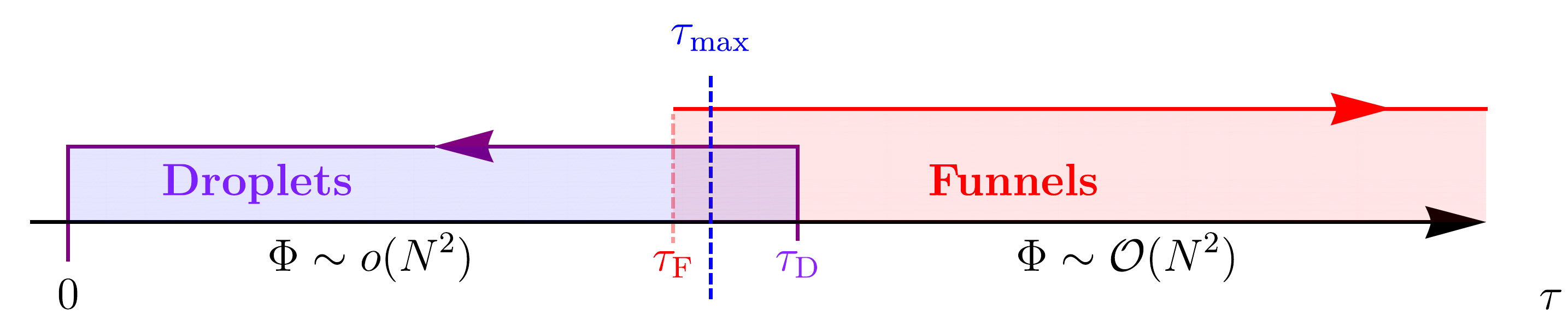}
\caption{Phase diagram of $\mathcal{N}=4$ SYM with gauge group $SU(N)$ and asymptotic temperature $T_{\infty}$, at large 't Hooft coupling and infinite $N$ on a Schwarzschild black hole background with temperature $T_{\mathrm{BH}}$. In this phase diagram $\tau\equiv T_{\infty}/T_{\mathrm{BH}}$.}
\label{fig:phase_diagram}
\end{figure}

We can now merge the results of \cite{Santos:2014yja} to the ones found in this manuscript, to infer a complete phase diagram for $\mathcal{N}=4$ SYM with gauge group $SU(N)$ and asymptotic temperature $T_{\infty}$, at large 't Hooft coupling and infinite $N$ on a Schwarzschild black hole background with temperature $T_{\mathrm{BH}}$. We attempt to draw such diagram in Fig.~\ref{fig:phase_diagram} as a function of $\tau\equiv T_{\infty}/T_{\mathrm{BH}}$. The black droplets of \cite{Santos:2014yja} were found to exist in the window $\tau\in[0,\tau_D]$, with $\tau_D\approx 0.93$. The solution with $\tau=0$, corresponding to the Unruh vacuum, was found previously in \cite{Figueras:2011va}. In the droplet phase, there is no $\mathcal{O}(N^2)$ Hawking radiation, and the corresponding state can be best understood in terms of a `jammed phase', though no underlying understanding of this phenomena exists to date on the field theory side. The funnel solutions we found in this manuscript exist at least in the range $\tau>\tau_{F}\approx0.557103$, though we expect such solutions to exist for even lower values of $\tau$, which we are unable to probe with current numerical techniques. Most notably, the Hawking flux $\Phi$  in the flowing funnel phase does not appear to be monotonic with decreasing $\tau$, reaching a maximum at $\tau=\tau_{\max}\approx0.698282$. Borrowing intuition from the Stefan-Boltzman law, this would suggest that the effective number of degrees of freedom in the field theory acquires a non-trivial temperature dependence.
We conjectured that flowing funnels with $\tau<\tau_{\max}$ are linearly unstable to the Gregory-Laflamme instability, and that the concomitant endpoint is one of the droplets found in \cite{Santos:2014yja}. Using results from \cite{Figueras:2011he}, one can give numerical evidence in favour of this conjecture, which will be presented elsewhere. It is clear that our results will largely be stable to corrections in the 't Hooft coupling, which manifest themselves as higher derivative corrections in the bulk. However, this is not the case for finite $N$ effects. At finite $N$, we expect Hawking radiation in the bulk, and thus a non-trivial flux of Hawking radiation even in the droplet phase. We stress, however, that this effect is subleading in $N$, perhaps being $\mathcal{O}(1)$, instead of the $\mathcal{O}(N^2)$ effect of the flowing funnel phase.

The most resounding mystery of this work still pertains the existence of the droplet phase \cite{Figueras:2011va,Santos:2014yja}, in which Hawking radiation seems trapped by the local geometry and does not escape to infinity, perhaps due to some form of local confinement yet to be explored on the quantum field theory side.

\subsection*{Acknowledgments}
J.~E.~S. thanks Sebastian~Fischetti, Juan~Maldacena, Don~Marolf, Benson~Way and Edward~Witten for many discussions and \'O.~J.~C.~Dias, B.~Ganchev and J.~F.~Melo for reading an earlier version of this manuscript. J.~E.~S. also thanks Toby~Crisford for many discussions and collaboration in the initial stages of this work. J.~E.~S. is supported in part by STFC grants PHY-1504541 and ST/P000681/1. J.~E.~S. also acknowledges support from a J.~Robert~Oppenheimer Visiting~Professorship. This work used the DIRAC Shared Memory Processing system at the University of Cambridge, operated by the COSMOS Project at the Department of Applied Mathematics and Theoretical Physics on behalf of the STFC DiRAC HPC Facility (www.dirac.ac.uk). This equipment was funded by BIS National E- infrastructure capital grant ST/J005673/1, STFC capital grant ST/H008586/1, and STFC DiRAC Operations grant ST/K00333X/1. DiRAC is part of the National e-Infrastructure.
\appendix
\section{Convergence tests}
\begin{figure}[h]
\centering
\includegraphics[width=\linewidth]{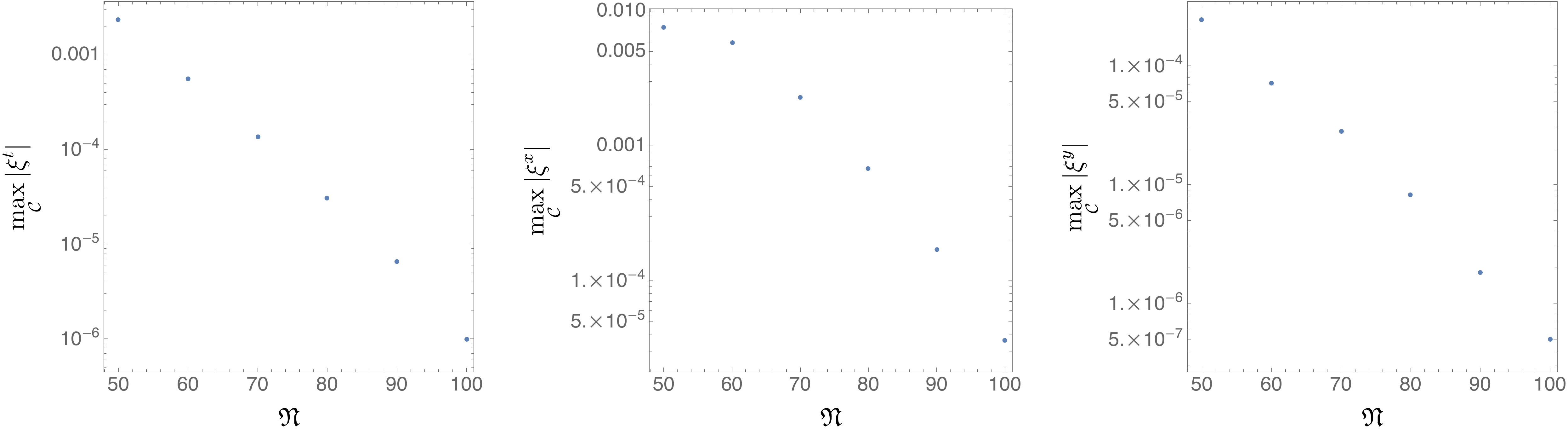}
\caption{Convergence properties of the non-vanishing components of DeTurck vector $\xi^a$, as a function of the number of grid points $\mathfrak{N}$. From left to right, we plot in a logarithmic scale $\max_{\mathcal{C}} |\xi^t|$, $\max_{\mathcal{C}} |\xi^x|$, $\max_{\mathcal{C}} |\xi^y|$, with $\mathcal{C}$ being the domain of outer communications. All plots were generated using $\tau= 0.665557$.}
\label{fig:convergenceturck}
\end{figure}

We monitored all the components of $\xi^a$ as a function of the number of grid points in the $x$ and $y$ directions, $\mathfrak{N}_x$ and $\mathfrak{N}_y$, respectively. For simplicity, we present results for $\mathfrak{N}_x=\mathfrak{N}_y=\mathfrak{N}$. Note that since the problem is mixed Elliptic-Hyperbolic, it does not suffice to monitor the norm $\xi^a \xi_a$, as $\xi^a$ can be null in certain regions of our integration domain. Independently of the temperature ratio $\tau\equiv T_{\infty}/T_{\mathrm{BH}}$, we found that the approach to the continuum limit was consistent with power law convergence, with the details of the convergence depending slightly on which component of $\xi$ one looks at. This convergence properties is to be expected, since non-analytic terms were identified in an expansion off the conformal boundary (see Eq.~(\ref{eq:expandcoord})). In Fig.~\ref{fig:convergenceturck} we show a logarithmic plot of $\max_{\mathcal{C}} |\xi^t|$ (left), $\max_{\mathcal{C}} |\xi^x|$ (middle) and $\max_{\mathcal{C}} |\xi^y|$ (right) as a function of $\mathfrak{N}$, where $\mathcal{C}$ stands for domain of outer communications. For $\tau= 0.665557$, we find $\max_{\mathcal{C}} |\xi^t|\propto \mathfrak{N}^{-12.22}$, $\max_{\mathcal{C}} |\xi^x|\propto \mathfrak{N}^{-9.89}$ and $\max_{\mathcal{C}} |\xi^y|\propto \mathfrak{N}^{-9.83}$.

\begin{figure}[h]
\centering
\includegraphics[width=.57\linewidth]{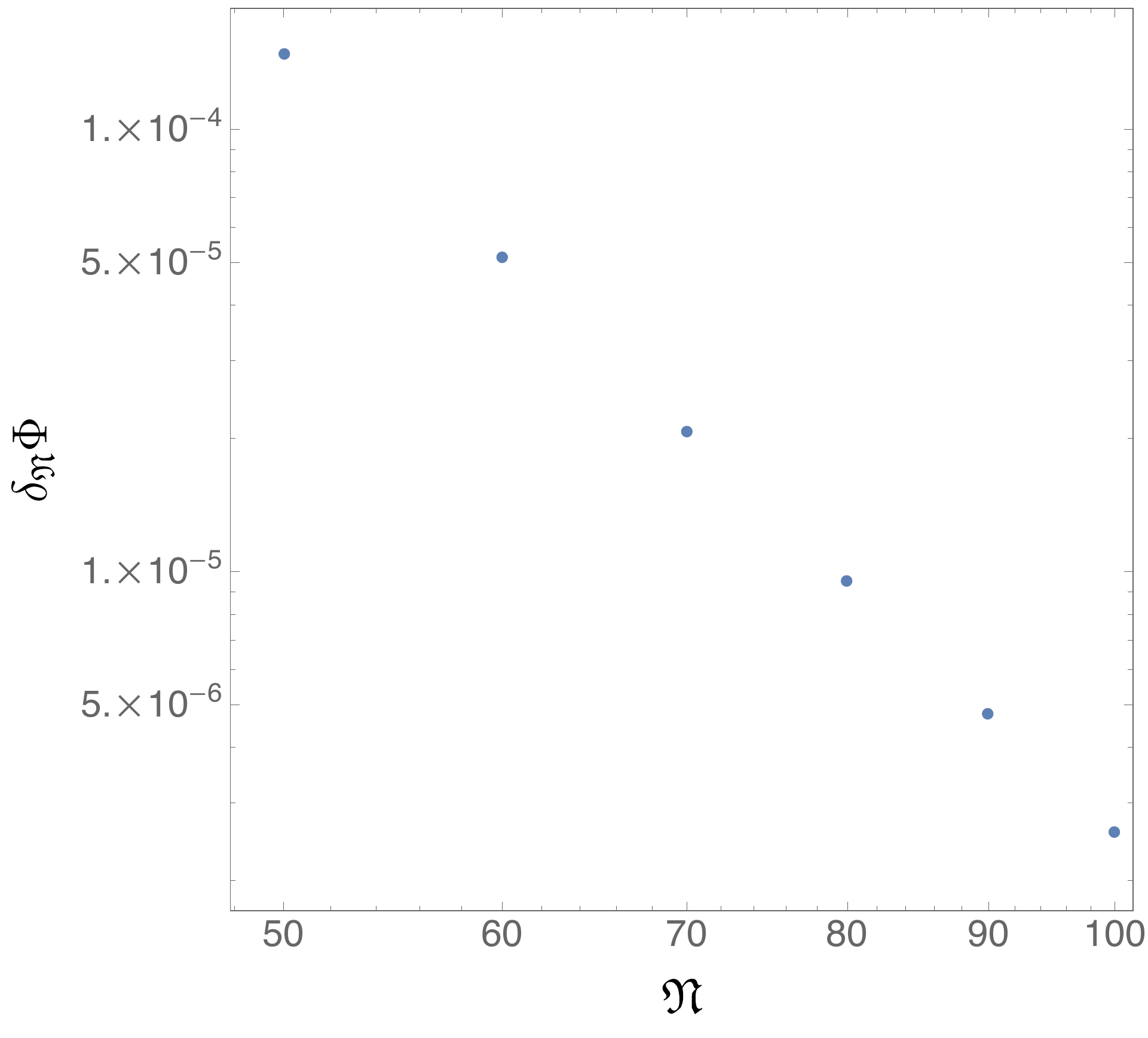}
\caption{ $\delta _{\mathfrak{N}}\Phi$ as a function of $\mathfrak{N}$ computed for fixed $\tau = 0.7293$. Convergence is now consistent with $\delta_{\mathfrak{N}}\Phi\propto \mathfrak{N}^{-5.85}$}
\label{fig:convergencealpha}
\end{figure}

Finally, we also investigated the convergence properties of $\Phi$, by defining
\begin{equation}
\delta _{\mathfrak{N}}\Phi\equiv \left|1-\frac{\Phi_{\mathfrak{N}}}{\Phi_{\mathfrak{N}+10}}\right|\,,
\end{equation}
where $\Phi_{\mathfrak{N}}$ stands for $\Phi$ computed on a mesh with $\mathfrak{N}_x=\mathfrak{N}_y=\mathfrak{N}$ grid points. The results can be seen in Fig.~\ref{fig:convergencealpha} for $\tau = 0.7293$ where we now find $\delta_{\mathfrak{N}}\Phi\propto \mathfrak{N}^{-5.85}$, which is very close to the non-analytic power reported in the expansion (\ref{eq:expandcoord}).
\bibliography{hawkingradiation}{}
\bibliographystyle{JHEP}
\end{document}